\documentclass[twocolumn,showpacs,preprintnumbers,amsmath,amssymb]{revtex4}

\usepackage{graphicx}

\begin{document}

\preprint{TQO-ITP-TUD/09-2008}

\title{Stochastic field equation for the canonical ensemble of a Bose gas}

\author{Sigmund Heller}
\affiliation{Institut f\"{u}r Theoretische Physik, 
Technische Universit\"at Dresden, D-01062 Dresden, Germany}
\author{Walter T. Strunz}
\affiliation{Institut f\"{u}r Theoretische Physik, 
Technische Universit\"at Dresden, D-01062 Dresden, Germany}

\date{\today}

\begin{abstract}
We present a novel norm preserving stochastic evolution equation for a Bose
field. Ensemble averages are quantum expectation values in the
canonical ensemble. This numerically very stable equation suppresses 
high-energy fluctuations exponentially, preventing cutoff 
problems to occur.
We present 3D simulations for an ideal gas in various trapping potentials
focussing on ground state occupation numbers and spatial correlation
functions for a wide range of temperatures above and below the critical
temperature. Although rigorously valid for non-interacting Bosons only,
we argue that weakly interacting Bose gases may also be amenable to
this approach, in the usual mean-field approximation.
\end{abstract}

\pacs{05.30.Jp, 67.85.-d, 02.50.Ey}
\maketitle

Ultracold quantum gases in traps are currently being investigated to a 
hitherto unknown precision and in a variety of circumstances 
\cite{Pet02,Pit03,Blo07}.
The fascinating experimental possibilities to manipulate relevant
parameters such as trap geometry, temperature, particle number, and even
interaction strength show that these gases are
ideal quantum systems to investigate and revisit fundamental concepts of
many particle and statistical physics.

In these experiments, after cooling, traps 
contain roughly a fixed number of particles. Thus, from a physical
point of view, a canonical ensemble is to be preferred over a grand canonical
one. For these finite systems, different predictions for fluctuations
(even in the thermodynamic limit \cite{Zif77}) call for a canonical 
description.

We present a stochastic evolution equation for a c-number field 
$\psi(x)$ such that quantum statistical expectation values in the 
canonical state can be replaced by an ensemble mean 
over these stochastic fields.
As applications, we focus on densities, ground state occupation numbers, 
and (spatial) correlation functions which have recently been measured in
impressive experiments \cite{Blo00,Hel03,Foe05,Oet05,Sch05}.

We stress two
crucial properties of our novel equation: first, the noise is
spatially correlated, preventing cutoff problems to occur.
Secondly, the equation is norm-preserving, reflecting the canonical
nature of our ensemble. Both these properties
ensure a very stable numerical solution such that full 3D problems 
may be tackled. Moreover, our equation may be implemented in
position space such that arbitrary 
trapping potentials may be treated without any difficulty, and, 
eventually, interactions may be taken into account.

While constructed for an ideal gas, we do strongly believe
that these positive features of our stochastic field equation (SFE)
will also be valuable for the interacting case.
There are a number of approaches that establish SFEs
for the grand canonical state of an interacting Bose gas in a trap 
\cite{Sto97,Dav01,Gar02,Bra05,Khe04}. We will comment on
these equations and their relation to our result 
towards the end of this work. 

We consider an ideal gas of $N$ particles in a trap
with single-particle Hamiltonian $H=\frac{p^2}{2m}+V(x)=
\sum\limits_k \epsilon_k |\epsilon_k \rangle\langle\epsilon_k |$
and eigenenergies $ \epsilon_k$. For the determination of 
$N$-particle mean values $\langle \ldots \rangle_N = $tr$(\ldots
\hat{\rho}_N )$ we start with the
canonical density operator
\begin{equation}\label{density}
\hat{\rho}_N=\frac{1}{Z_N}e^{-\hat{H}/kT}\hat{P}_N
\end{equation}
in second quantization with the corresponding energy
$\hat{H}= 
\sum\limits_k \epsilon_k\hat{a}^{\dagger}_k\hat{a}_k$,
the canonical partition function $Z_N$, 
and the projector 
$\hat{P}_N=\sum\limits_{\sum n_k=N}|\{n_k\}\rangle\langle\{ n_k \}|$ 
onto the $N$-particle subspace. As usual, the number states are
$|\{n_k\}\rangle=
\prod\limits_k(\hat{a}^{\dagger}_k)^{n_k}/\sqrt{n_k!}|0\rangle$, 
where $n_k$ is the occupation number of the k-th eigenstate 
$|\epsilon_k\rangle$.

With the notation $\psi(x) = \langle x| \psi\rangle$ 
the desired SFE takes the form
of a nonlinear, norm preserving stochastic Schr\"odinger
equation, here in Stratonovich calculus \cite{Gar83} 
\begin{eqnarray}\label{quantumequation}
d |\psi\rangle=-\frac{1}{\hbar}
\left((\Lambda+i)H-\Lambda\frac{\langle\psi|H|\psi\rangle}
{\langle\psi|\widehat{kT}|\psi\rangle}\widehat{kT}\right)
|\psi\rangle dt\nonumber\\
+\sqrt{\frac{2\Lambda}{\hbar}}\left(\sqrt{\widehat{kT}}
|d\xi\rangle-\frac{\langle\psi|
\sqrt{\widehat{kT}}|d\xi\rangle}{\langle\psi|\widehat{kT}|\psi\rangle}
\widehat{kT} |\psi\rangle\right),
\end{eqnarray}
which is the central result of this paper.
In equ. (\ref{quantumequation}), $\Lambda$ is a phenomenological 
damping parameter.
Reflecting a fluctuation-dissipation-relation, it
appears both, in the damping term and, as a square root, in the
fluctuations.
Crucially, we introduce an {\it operator}
$\widehat{kT}=\frac{H}{e^{H/kT}-1}$
that depends on the real temperature $kT$, obviously a reference to
the Bose occupation number.
The complex random
field $d\xi(x,t) = \langle x| d\xi(t)\rangle$ represents
white noise with correlations 
$\langle d\xi(t)|x\rangle\langle x'|d\xi(t)\rangle=\delta(x-x')dt$.
Note, however, that the white noise is always acted upon by the operator 
$\sqrt{\widehat{kT}}$. For energies $H\ll kT$ the operator
$\widehat{kT}$ acts simply as the multiplication with the thermal energy
$kT$.  However, for energies 
$H \gg kT$, the fluctuations are exponentially suppressed,
which is a crucial feature of our novel equation to which 
we will come back towards the end of the paper.
Another way of looking at this is that proper quantum statistics 
leads to spatially correlated noise 
preventing the occurrence of arbitrarily high momenta \cite{hellerstrunz02}.
 
We found equation (\ref{quantumequation}) by starting with the
Glauber-Sudarshan P-representation \cite{Sch01}
of the Gaussian exponential in (\ref{density}),
\begin{equation}\label{p-function}
\frac{e^{-\hat{H}/kT}}{Z} = 
\int\frac{d^2\psi}{\pi}P(\psi^{\ast},\psi)|\{\psi\}\rangle\langle\{\psi\}|.
\end{equation}
Coherent states
$|\{\psi\}\rangle= |\psi_0\rangle|\psi_1\rangle\cdots|\psi_k\rangle\cdots$
are used for all modes and 
$P(\psi^{\ast},\psi)=\prod\limits_k(e^{\epsilon_k/kT}-1)\cdot
\exp\left(-\sum\limits_k|\psi_k|^2(e^{\epsilon_k/kT}-1)
\right)$ (see \cite{Sch01}). The fact that
$\langle\{ \psi\} |{\hat{P}}_N| \{\psi\}\rangle
= \langle\psi|\psi\rangle^N 
\exp\{-\langle\psi|\psi\rangle\}/N!$ with
$\langle\psi|\psi\rangle = \sum_k|\psi_k|^2$ (see \cite{Mol68}),
allows us to express (normally-ordered)
quantum correlation functions as
phase space integrals, for instance
\begin{equation}\label{firstcorrelations}
\langle \hat{\psi}^{\dagger}(x)\hat{\psi}(x')\rangle_N =
\frac{1}{C_N}\int\frac{d^2\psi}{\pi}\psi^{\ast}(x)\psi(x')
W_{N-1}(\psi^{\ast},\psi).
\end{equation} 
The weight functions are given by
$W_N(\psi^{\ast},\psi)=\frac{1}{N!}\,\langle\psi|\psi\rangle^{N}\,
e^{-\langle\psi|\psi\rangle}\,P(\psi^{\ast},\psi)$
with the normalization constant 
$C_N=\int\frac{d^2\psi}{\pi}W_{N}(\psi^{\ast},\psi)$. Note
that second (or higher) order correlations have to be determined
using $W_{N-2}$ (or lower index) in expression (\ref{firstcorrelations}), 
while the $C_N$ remains.

We stress that it is necessary to distinguish carefully
between the norm 
${\cal N}=\langle\psi|\psi\rangle$ of the stochastic field evolving
according to
equ. (\ref{quantumequation}) -- which remains constant for all times --
and the particle number $N$.
Apart from the solution of our (rigorous) SFE (\ref{quantumequation}),
the exact stochastic simulation of the weight function $W_N$ 
requires a distribution of values for ${\cal N}$. It turns out,
however, that for particle numbers 
much larger than one the norm distribution becomes narrow enough
so that for all the temperatures
and observables of interest in this paper a 
simulation with a single norm ${\cal N}$ is sufficient
(see \cite{hellerstrunz02}). 
Still, there is a surprising subtlety: The distribution of the 
norm ${\cal N}$ depends on the absolute value of the ground state 
energy $E_0$. It is for $E_0=0$ only that we have to chose
$\langle {\cal N} \rangle = N$. The liberty to choose other values for $E_0$
(and other norms ${\cal N}$, accordingly) 
can be used with benefit to achieve faster convergence in 
the numerical implementation (see \cite{hellerstrunz02}). 

Two further remarks are called for. First, being a highly nonlinear 
equation, we are not surprised to find that it is possible
to replace the ensemble mean by a time average
$\langle\hat{\psi}^{\dagger}(x)\hat{\psi}(x')\rangle_N=
\lim\limits_{t\rightarrow\infty}\frac{1}{t}\int\limits_0^{t}ds\,
\psi^{\ast}(x,s)\psi(x',s)$ over a single realization $\psi(x,s)$.
Secondly, we chose to propagate with
the term $(\Lambda+i)$ in our equ. (\ref{quantumequation})
such that for $\Lambda=0$ the remaining complex unit ``$i$''
describes ``real'' dynamics. In this way we can simulate the transition 
from a non-equilibrium to an equilibrium state in a
phenomenological manner -- see recent experiments \cite{Rit07}.

\begin{figure}[t]
  \centering
    \includegraphics[width=6.5cm]{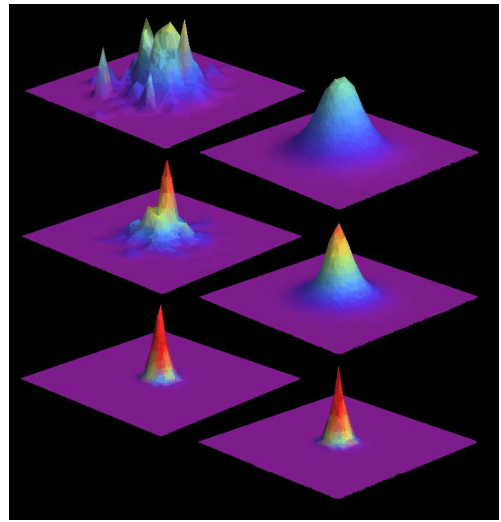}
  \caption{Momentum distribution of a Bose gas of $200$ 
 particles in a potential $V(x,y,z)=x^4+y^4+z^4$ during phase 
 transition. On the left hand side we show single realizations of 
 our stochastic 
 field equation (\ref{quantumequation}), on the right hand side 
 long-time averages over 15000 time steps.
  Top: an almost ``classical'' distribution for $T>T_c$;
  Middle: a peak develops for $T\approx T_c$;
  Bottom: most particles are condensed for $T<T_c$.}
  \label{b1}
\end{figure}

\begin{figure}[t]
  \centering
    \includegraphics[width=6cm]{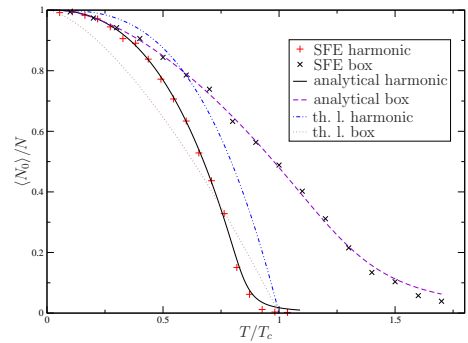}
 \caption{Ground state occupation as a function of temperature for 
 an ideal 3D Bose gas of 200 particles in a harmonic isotropic trap 
 (plus signs) and for 100 particles in a box potential (crosses).
 We compare with an analytical calculation for the harmonic 
 (in a low temperature approximation) \cite{Koc00} (full line)
 and the box potential \cite{Gla07} (dashed line). 
 We also display results for the thermodynamic limit
 (harmonic: dashed-dotted line, box: dotted line).}
  \label{b2}
\end{figure}

We now turn to applications. While a gas in a box is best treated in
momentum space, a general implementation of equ. (\ref{quantumequation}) in
position space is advantageous, since it can be adjusted easily to any
trapping potential (and in a next step mean field atomic interactions may 
be included -- see later). However, in position space, the 
generation 
of the correlated noise $\langle x|\sqrt{\widehat{kT}}|d\xi\rangle$ is 
cumbersome.
For the simulations presented here we use a Wigner-Weyl representation 
\cite{Sch01} of the operator $\widehat{kT}$ and consider only terms of lowest
order in $\hbar$.
As the examples below show, this approximation is
legitimate (for more details, see \cite{hellerstrunz02}). 

The functioning of our equation is visualized in Fig. \ref{b1}.
We simulate a 3D Bose gas containing 200 particles, trapped in a
quartic potential $V(x,y,z)=x^4+y^4+z^4$ and
determine the momentum densities $n(p_x,p_y)$ 
(the third momentum $p_z$ is integrated over). 
Three different temperatures are chosen:
above (top), at (middle), and below (bottom) 
the critical temperature for Bose-Einstein condensation. 
While on the left hand side we display a single realization 
of eq. (\ref{quantumequation}) after a certain
propagation time, the right hand side shows time averages 
over 15000 time steps. Obviously, on average we obtain the typical 
pictures for the transition to a Bose-Einstein condensate.

The true quality of our equation is to be verified by calculating 
various characteristic quantities. As shown in Fig. \ref{b1}, 
the numerical code in position representation allows us to treat the 
Bose gas in any trapping potential. In order to make contact to
previous results for the canonical ensemble, however, 
we restrict ourselves in the following to a 3D box and a 3D
harmonic oscillator potential. More general considerations will be 
published elsewhere \cite{hellerstrunz02}.

First we show the ground state
occupation in Fig. \ref{b2}. The results for the harmonic
oscillator (with $N=200$ particles, plus signs) are 
obtained by propagating equ. (\ref{quantumequation}) on a position grid.
No use is made of the known spectrum and eigenfunctions.
We compare with an analytical 
approximation (full line) \cite{Koc00},
which is known to be in good agreement with exact results, and the 
thermodynamic limit (dashed-dotted line). 
Next we show results for a 3D box
($N=100$ particles, crosses) obtained from our SFE
(here computed in momentum space) compared with an (approximate) result 
based on a path integral approach \cite{Gla07} (dashed line) and find very good
agreement. Temperature is scaled to the critical temperature of
the thermodynamic limit \cite{Pet02,Pit03}. Note that finite size effects are
very significant for the box as seen when comparing with the
thermodynamic limit (dotted line).

\begin{figure}[t]
  \includegraphics[width=10cm]{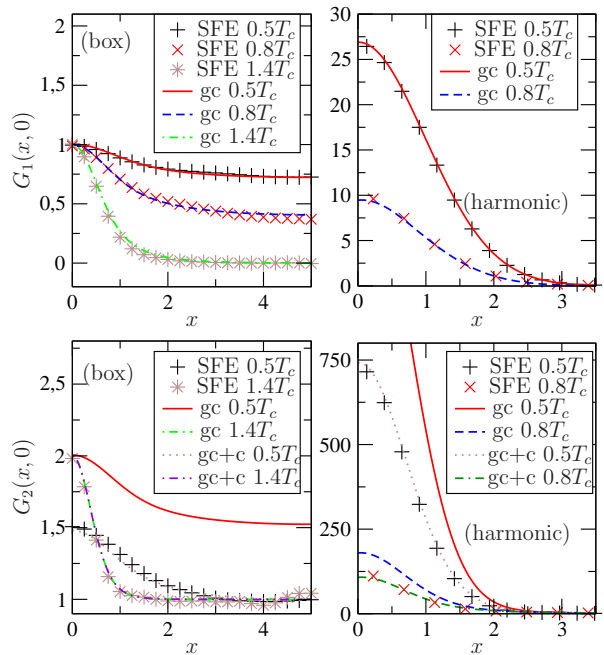}
  \caption{First and second order spatial correlations $G_1(x,0)$ (top) and
  $G_2(x,0)$ (bottom) for a Bose gas in a 3D box potential with periodic 
  boundary 
  conditions (left) and a 3D harmonic oscillator potential (right) for 
  different temperatures. Results from the SFE
  (plus signs, crosses, stars) are compared to a direct calculation 
  based on the theory of \cite{Nar99}: 
  a grand-canonical description (``gc'', solid, dashed and dashed-dotted 
  lines) fails for $G_2$, while a corrected theory (``gc+c'' dashed-dotted and
  dotted lines) shows good agreement.
}
  \label{b3}
\end{figure}

Next we determine spatial correlation functions of first
$G_1(x,x')=\langle\hat{\psi}^{\dagger}(x)\hat{\psi}(x')\rangle_N$
and second order 
$G_2(x,x')=\langle\hat{\psi}^{\dagger}(x)\hat{\psi}^{\dagger}(x')
\hat{\psi}(x)\hat{\psi}(x')\rangle_N$
above and below the critical temperature.
$G_1$ can be measured through interference experiments (see
\cite{Blo00}). $G_2$ and related quantities
have been measured more recently in impressive
experiments \cite{Hel03,Foe05,Oet05,Sch05}.
For first order correlations, the 
canonical results are close to the grand canonical values. 
For second order correlations 
and temperatures below the critical temperature, however,
large deviations appear and the results of the canonical ensemble 
may be obtained approximately with the help of
condensate corrections to the grand canonical calculation \cite{Nar99}.
In Fig. \ref{b3},
we show $G_1(x,0)$ (top) and $G_2(x,0)$ (bottom)
obtained from the SFE (\ref{quantumequation}); 
here for a 3D Bose gas of $1000$ particles in a box
with periodic boundary conditions (left hand side) and for
a gas of $200$ particles in an isotropic harmonic oscillator (right hand side).
Our values (plus signs, crosses, stars) are compared with a direct 
grand canonical calculation (``gc'' in Fig. \ref{b3}) and the
corrected description  (``gc+c'' in Fig. \ref{b3})
based on the theory of \cite{Nar99}.
These corrected second order correlation functions are in good
agreement with our exact canonical calculations.

We see the main achievement of this paper in the fact that
we established a numerically very robust (exact) SFE
for the canonical state of an ideal Bose gas, amenable to
an efficient numerical solution in full 3D for arbitrary trapping
potentials. Still, in the light of the wealth of activities involving
ultracold quantum gases,
it is certainly of great importance to also investigate the 
interacting case. Let us therefore relate equ. (\ref{quantumequation})
to previous stochastic equations constructed for the
grand canonical ensemble of an interacting Bose 
gas \cite{Sto97,Dav01,Gar02,Bra05,Khe04}.
A good survey over these different findings, their relations and their
limitations is given in \cite{Bra05}. 
Most notably, in several of these approaches
\cite{Sto97,Dav01,Gar02,Bra05}, termed
``classical field methods'' in \cite{Bra05} (and see also
\cite{Hoh77}), due to ultraviolet problems, lowly occupied states must 
be cut off \cite{Dav01} or
treated in a different formalism \cite{Gar02,Bra05}.

Let us now turn to our SFE (\ref{quantumequation}):
omitting the non-linear terms and substituting both
$H\rightarrow {\tilde H} = H-\mu{\hat N}$
and $\widehat{kT} \rightarrow
\widetilde{kT} = {\tilde H}/(e^{{\tilde H}/kT} -1)$
($\mu$ the chemical potential and 
${\hat N}$ the number operator), equ.
(\ref{quantumequation}) reduces to 
\begin{equation}\label{grandcanonical}
d |\psi\rangle=-(\Lambda+i)\tilde{H}|\psi\rangle+
\sqrt{2\Lambda\widetilde{kT}}\;|d\xi\rangle
\end{equation}
of which one can easily show that it indeed provides proper
grand canonical ensemble averages \cite{hellerstrunz03}.
We hasten to stress that the canonical 
equation (\ref{quantumequation}) is not merely a
normalized version of equ.(\ref{grandcanonical}). 
As discussed before, in (\ref{grandcanonical})
the operator $\widetilde{kT}$ incorporates 
a natural high energy noise cutoff induced by proper quantum 
statistics. No further care is required.

Describing a grand canonical ensemble,
equ. (\ref{grandcanonical})
is the link to establish a connection to
the ``classical field methods'' mentioned above for an interacting gas:
First, the operator $\widetilde{kT}$ appears as the simple
temperature $kT$ in those approaches, requiring a cutoff.
More importantly, the interaction may be taken into account
by a mean field contribution
${\tilde H}\rightarrow {\tilde H}+g|\psi(x)|^2$, where $g=4\pi a \hbar^2/m$
is the interaction parameter and $a$ the s-wave scattering length.

After these considerations it appears more than tempting to use
equation (\ref{quantumequation}) for a canonical ensemble even in the 
case of a weakly interacting gas, with
$V(x)\rightarrow V(x)+g|\psi(x)|^2$. As argued,
the resulting equation is free from ultraviolet problems
and coincides (in the grand canonical case) with previous 
(``classical'') equations
(including interaction and a high-energy cutoff).
Moreover, it reduces to the (imaginary-time) 
Gross-Pitaevskii equation for $kT\rightarrow 0$, describing a 
pure condensate (with the given particle number $N$).

Let us briefly summarize our result: we present a norm-preserving
stochastic field equation for the canonical state of a Bose gas,
describing experiments with a finite number of atoms in an arbitrary 
trap. Being driven by spatially correlated noise,
cut-off issues do not appear; the equation is numerically very stable.
We stress that it is valid for arbitrary temperatures: 
for $T\gg T_c$ it provides a wave description of  
a classical gas of massive particles. This is very much in the spirit of
the way we think of light emerging from a light bulb as being
composed of incoherent wave trains. We are able to determine
important quantities like spatial correlation functions and
occupation numbers as a function of temperature in arbitrary traps. 
Finally, we relate our equation
to stochastic Gross-Pitaevskii equations that exist for interacting
gases in the grand canonical ensemble, raising
expectations that the new equation should be applicable to weakly 
interacting Bose gases as well.

We are grateful for inspiring discussions with Markus Oberthaler 
and Thimo Grotz. S.~H. acknowledges support by the International Max Planck 
Research School for Dynamical Processes in Atoms, Molecules and Solids,
Dresden.

\end{document}